\documentclass{aa}  
\usepackage{graphics}
\usepackage{txfonts}
\def\kms{km~s$^{-1}$}
\def\teff{${\rm T_{eff}}$}
\def\na{${\rm [Na/Fe]_{NLTE}}$}
\def\ali{${\rm A(Li)_{NLTE}}$}

\begin{document} 
  \title{The discovery of a Li-Na-rich giant star in Omega Centauri:\\
formed from the pure ejecta of super-AGB stars?
   \thanks{Based on observations collected at the ESO-VLT under programs 096.D-0728.}}

\author{A. Mucciarelli\inst{1,2}, L. Monaco\inst{3}, P. Bonifacio\inst{4}, M. Salaris\inst{5}, 
        X. Fu\inst{1,2}, S. Villanova\inst{6}}
\offprints{A. Mucciarelli}
\institute{
Dipartimento di Fisica e Astronomia, Universit\`a degli Studi di Bologna, Via Gobetti 93/2, I-40129 Bologna, Italy;
\and
INAF - Osservatorio di Astrofisica e Scienza dello Spazio di Bologna, Via Gobetti 93/3, I-40129 Bologna, Italy;
\and
Departamento de Ciencias Fisicas, Universidad Andres Bello, Fernandez Concha 700, Las Condes, Santiago, Chile
\and
GEPI, Observatoire de Paris, Universit{\'e} PSL, CNRS, Place Jules
Janssen, 92195 Meudon, France
\and
Astrophysics Research Institute, Liverpool John Moores University, 146 Brownlow Hill, Liverpool L3 5RF, United Kingdom 
\and
Departamento de Astronomia, Casilla 160-C, Universidad de Concepcion, Concepcion, Chile}

\authorrunning{A. Mucciarelli et al.}
\titlerunning{Li-rich stars in Omega Centauri}


\abstract{
We report the discovery of two Li-rich giant stars (fainter than the red giant branch bump) in the stellar 
system Omega Centauri using GIRAFFE-FLAMES spectra. These two stars have A(Li)=1.65 and 2.40 dex and they belong to the 
main population of the system ([Fe/H]=--1.70 and --1.82, respectively). The most Li-rich of them (\#25664) has [Na/Fe]=+0.87 dex that is $\sim$0.5 dex higher
than those measured in the most Na-rich stars of Omega Centauri of similar metallicity.
The chemical abundances of Li and Na in \#25664 
can be qualitatively explained by deep extra mixing efficient 
within the star during its RGB evolution or by super-asymptotic 
giant branch (AGB) stars with masses between $\sim$7 and 8 $M_{\odot}$.
In the latter scenario, this Li-Na-rich star could be formed from the pure ejecta of super-AGB stars before the dilution with pristine material occurs, or, alternatively,
be part of a binary system and experienced mass transfer from the
companion when this latter evolved through the super-AGB phase. 
In both these cases, the chemical composition 
of this unique object could allow to look for the first time
at the chemical composition 
of the gas processed in the interior of super-AGB stars.}


\maketitle

%

\section{Introduction}
\label{intro}

The lithium abundance -A(Li)\footnote{A(Li)=$\log{\frac{n_{Li}}{n_H}}$+12.}- in globular clusters (GCs) remains an unsolved and fascinating riddle. 
These stellar systems are known to have multiple populations (MPs), 
characterised by significant star-to-star variations in the abundances of elements involved 
in proton-capture reactions, i.e. He, C, N, O, Na, Mg, Al, often structured in coherent 
patterns. Among these patterns, the most evident is the Na-O anticorrelation, observed in all the
old GCs \citep{carretta09,mucciarelli09}, with the only exceptions so far 
of Ruprechet 106 \citep{villanova13} and E3 \citep{monaco18}. 
According to their Na and O abundances, GC stars  are roughly classified in 
first (1P) and second (2P) populations, indicating stars without or with chemical 
signatures from proton-capture reactions, respectively.
These reactions occur at temperatures exceeding $10^7$ K, in stellar layers  having no
Li, because it is expected to be totally destroyed at temperatures of a few times $10^6$ K. 
For this reason, 2P stars should be Li-free or exhibit a significant depletion of A(Li) 
with respect to 1P stars. Surprisingly, no relevant difference in A(Li) is found in 1P and 2P stars, 
with some GCs showing only hints of a Li-Na anticorrelation 
or a larger scatter in A(Li) in 2P stars with respect to 1P stars
\citep{pasquini05,lind09,gh09,monaco12,dobrov14} 
and other GCs with a remarkably similar Li content among the stars \citep{boni02,dorazi14,dorazi15a}.
Theoretical models proposed to explain the formation of MPs suggest different
polluters in order to produce the observed Na-O anticorrelation, the
most popular ones being fast-rotating massive \citep{decressin07} and asymptotic giant branch (AGB) stars \citep{dercole08}. 
The former are not able to produce fresh Li, while the latter (especially the most massive ones, the so-called super-AGB stars) 
can produce new Li through the Cameron-Fowler mechanism \citep{fowler71}. 

In this framework, the stellar system Omega Centauri (NGC~5139) is a special case, 
showing a large [Fe/H] spread 
\citep[at variance with GCs that are homogeneous in their Fe content, see e.g.][]{willman12} 
but also an extended Na-O anticorrelation \citep{johnson10,marino11}. 
Also, Omega Centauri exhibits a clear and well-defined 
Li-Na anticorrelation \citep[][hereafter M2018]{mu18}. 
In this Letter we report the discovery of two giant stars in Omega Centauri 
showing a significant enhancement of A(Li) with respect to other stars of the system. 
One of these stars is also significantly enriched in Na.


\section{Observations and data analysis}
\label{obs}

In M2018 we presented Li, Na and Fe abundances 
of 199 lower red giant branch (LRGB, RGB stars fainter than the RGB bump magnitude level and having already completed 
the first dredge-up) stars, observed with the multi-object
high-resolution spectrograph FLAMES-GIRAFFE \citep{pasquini00} mounted at the ESO Very Large Telescope 
under the program 096.D-0728 (PI: Mucciarelli).  We secured one exposure with the setup HR12
($\Delta\lambda$=5821-6146 \AA , R$\sim$20000), 
two with HR13 ($\Delta\lambda$=6120-6405 \AA , R$\sim$26000) 
and three with HR15N ($\Delta\lambda$=6470-6790 \AA , R$\sim$19000).
The selected GIRAFFE gratings allow to measure the resonance Li line at 6708 \AA\ , 
the Na D doublet at 5890-5896 \AA\ and some tens 
of Fe~I lines. 
In this sample, we identified two additional LRGB stars
(namely \#25664 and \#126107 from the Bellini et al. 2009 catalogue)
 with a strong enhancement of A(Li). 

Here we briefly summarise the approach used for the chemical analysis, referring the reader 
to M2018 for a detailed description.
Effective temperatures (\teff) and surface gravities (log~g) have been derived 
from photometry. We adopted the \citet{alonso99} color-\teff\ transformations, 
using the ${\rm (B-V)}_0$, ${\rm (V-I)}_0$ and ${\rm (V-K_{s})}_0$ broad-band colours from the optical photometry 
by \citet{bellini09} and the 2MASS near-infrared database \citep{skrutskie}. 
The employed colour excess and distance modulus are E(B-V)=0.12 mag \citep{harris10} 
and $(m-M)_0$=13.70 mag \citep{bellazzini04}.
Microturbulent velocities have been derived by minimising the trend between the line strength 
and the abundances of the Fe~I lines.

Abundances of Fe have been derived from the measured equivalent widths 
using the {\tt GALA} code \citep{m13g}, while the equivalent widths have been 
measured with the {\tt DAOSPEC} code \citep{stetson08} 
managed through the wrapper {\tt 4DAO} \citep{4dao}.
The abundances of Li and Na have been obtained through a
$\chi^2$-minimisation between the observed and synthetic spectra 
\citep[the latter calculated with the {\tt SYNTHE} code,][]{kurucz05}. 
These abundances have been corrected for non local thermodynamical equilibrium 
(NLTE) using the corrections by \citet{lind08} and \citet{lind11} for Li and Na, respectively.
Information about the two stars is listed in Table~\ref{tab1}.

\section{Results}

Both stars are members of Omega Centauri, as confirmed by their radial velocities (see Table 1) and  
proper motions \citep{bellini09,gaia18}. According to their iron content 
([Fe/H]=--1.82 and --1.70 dex for \#25664 and \#126107, respectively), the two stars belong 
to the main (metal-poor) population of the system (see Fig.~1 in M2018). 

\begin{figure}
\includegraphics[width=\columnwidth]{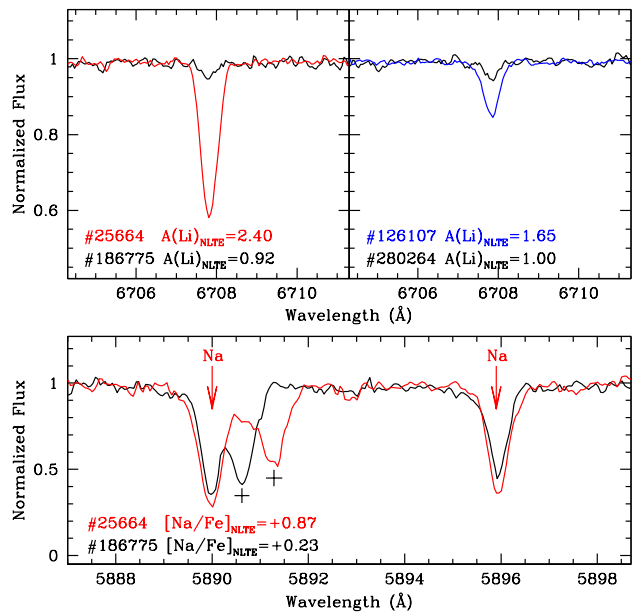}
\caption{The resonance Li line in \#25664 (left upper panel, red line) and \#126107 (right upper panel, blue line) 
both compared with stars of Omega Centauri of similar metallicity and atmospheric parameters (M2018) .
The lower panel shows the comparison between the \ion{Na}{i} D lines in \#25664 and in the comparison star. 
Plus symbols indicate the interstellar Na D2 lines.}
\label{spec}
\end{figure}

The two stars exhibit A(Li) larger than those measured 
in the other LRGB stars of Omega Centauri.
The A(Li) distribution of the LRGB stars in this system is peaked at A(Li)$\sim$0.9-1 dex
with a significant ($\sim$30\%) fraction of stars with low Li abundances (down to $\sim$0.4-0.5 dex). 
The fit of the resonance Li line at 6708 \AA\ provides abundances of 
\ali=~2.4 dex for \#25664 and \ali=~1.65 dex for \#126107.
The upper panels of Fig.~\ref{spec} show the Li line at 6708 \AA\ for the two stars 
in comparison with the (weaker) Li lines observed in two Omega Centauri stars with similar atmospheric parameters and metallicity 
(namely \#186775 and \#280264), and belonging to the main component 
of the A(Li) distribution (A(Li)$\sim$0.9-1 dex).

\begin{figure}
\includegraphics[width=\columnwidth]{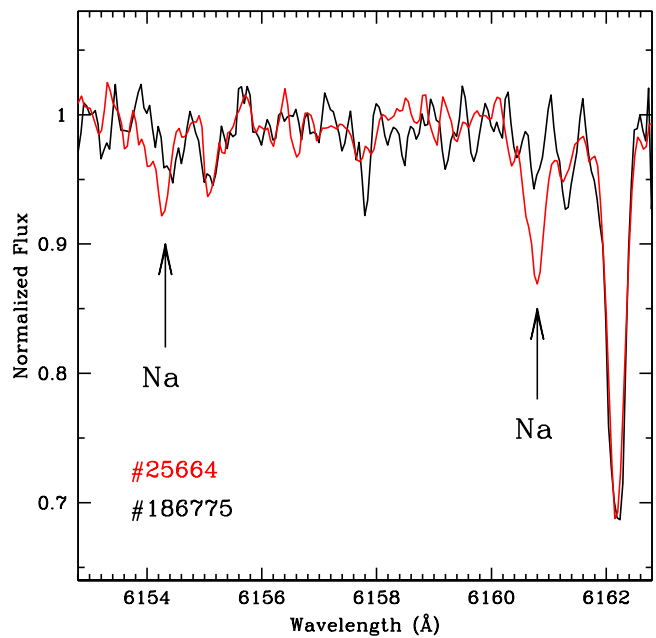}
\caption{The \ion{Na}{i} doublet at 6154-6160 \AA\ in \#25664 (red line) compared with the spectrum 
of the star \#186775 (black line).}
\label{spec2}
\end{figure}

Concerning Na, we derived \na=+0.87 and +0.14 for \#25664 and \#126107, respectively.
The lower panel of Fig.~\ref{spec} compares the strength of the \ion{Na}{i}~D lines in \#25664 and 
in the comparison star \#186775 (that is one of the most Na-rich stars among the metal-poor ones 
of Omega Centauri).
As a sanity check we also used the weak  \ion{Na}{i}  doublet at 6154-6160 \AA\ , 
available in the HR13 setup. This doublet provides  
\na=+1.07 and +0.08 for \#25664 and \#126107, respectively, 
confirming the extreme Na enhancement obtained from the 
Na D lines for the first star. 
In Fig.~\ref{spec2} we compare the spectra of 
\#25664 and of the comparison star \#186775 
(the same shown in Fig.~\ref{spec}), around the \ion{Na}{i} 
doublet at 6154-6160 \AA\ . 
These \ion{Na}{i} lines are located on the 
linear part of the curve of growth and they are
less sensitive to velocity fields and 
saturation effects with respect to the \ion{Na}{i} D lines. 
Because the two sets of lines 
provide compatible results, in the following we will 
refer to abundances derived from \ion{Na}{i} D lines to 
compare the abundances of the two Li-rich stars with those of the other stars of Omega Centauri discussed in M2018 
(in fact, for most of the stars analysed in M2018 the \ion{Na}{i} lines at 6154-6160 \AA\ 
are too weak and provide upper limits only).

\begin{figure}
\includegraphics[width=\columnwidth]{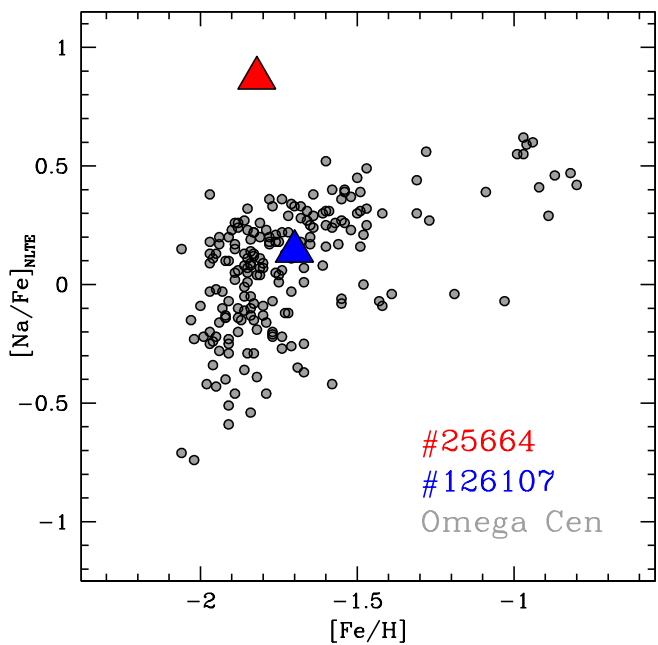}
\caption{Behaviour of \na\ as a function of [Fe/H] for the LRGB stars 
of Omega Centauri (grey circles, M2018). The red and blue triangles indicate
the position of \#25664 and \#126107, respectively. }
\label{nafe}
\end{figure}

Fig.~\ref{nafe} shows the behaviour of \na\ as a function of [Fe/H] for the LRGB stars of Omega Centauri 
(grey circles, M2018) with marked as red and blue triangles the two Li-rich stars. 
The metal-poor stars of Omega Centauri span a large range ($\sim$1 dex) in Na abundance, 
similar to the [Na/Fe] range measured in the old GCs studied so far \citep[see e.g.][]{carretta09}.
While \#126107 has a \na\ compatible with the Na distribution of stars with similar 
[Fe/H] (and it can be considered as a 2P star), 
\#25664 exhibits an extraordinary enhancement of Na, $\sim$0.5 dex larger than that measured 
in the most Na-rich Omega Centauri stars 
with similar [Fe/H] (see Fig.~\ref{nafe}). A very few GC stars with [Na/Fe] significantly larger than 
the most Na-rich stars in the parent cluster have been discovered so far, namely in Omega Centauri \citep{marino11}, 
NGC~2808 \citep{carretta06} and M62 \citep{lapenna15}.
All these stars are brighter than the RGB bump level but no measure of Li is available.

\begin{figure}
\includegraphics[width=\columnwidth]{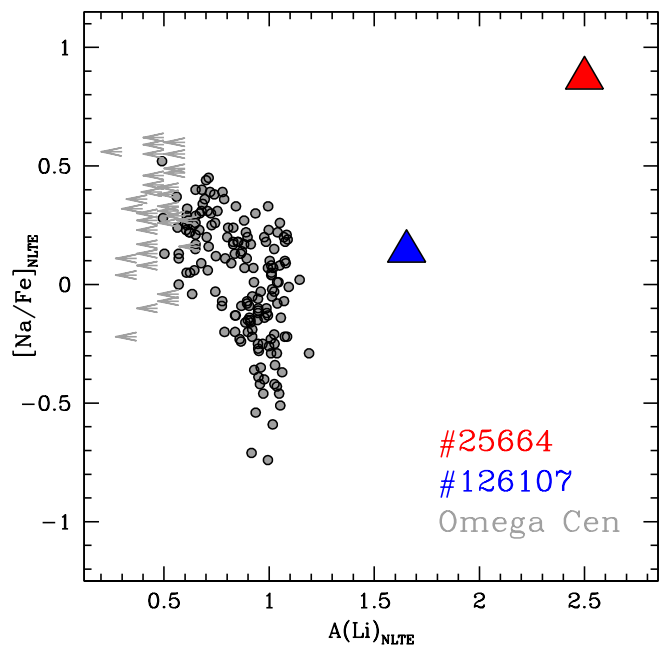}
\caption{Behaviour of \na\ as a function of \ali\ for the LRGB  
and the two Li-rich stars of Omega Centauri (same symbols of Fig.~\ref{nafe}).}
\label{nali}
\end{figure}

Fig.\ref{nali} shows the position of the two Li-rich stars in the \na-\ali\ plane. The 
LRGB stars of the system define a clear Li-Na anticorrelation, while the two target stars 
lie outside the mean locus defined by the other stars.

\begin{table*}
  \begin{center}
  \caption{Main information about the two Li-rich stars in Omega Centauri}
  \label{tab1}
  \begin{tabular}{lccccccccc}
  \hline
     ID  & RA  &Dec   &  RV    &${\rm T_{eff}}$ &  log~g   &  ${\rm v_t}$ &  [Fe/H]  &  \ali  & \na   \\
         &  (J2000)      &  (J2000)       & (\kms) &	(K)	    &	       &  (\kms)      &  (dex)   &  (dex) & (dex) \\  
\hline
    25664  &  201.751755  & --47.654579    &  +227.8$\pm$0.2  & 4958 & 2.37  & 1.5  &  --1.82$\pm$0.07 &  +2.40$\pm$0.06 &  +0.87$\pm$0.07  \\  
   126107  &  201.822311  & --47.533607    &  +258.7$\pm$0.1  & 4914 & 2.27  & 1.3  &  --1.70$\pm$0.06 &  +1.65$\pm$0.06 &  +0.14$\pm$0.06  \\  
\hline
\end{tabular} 
\end{center}
\end{table*}


\section{Discussion}

\begin{figure}
\includegraphics[width=\columnwidth]{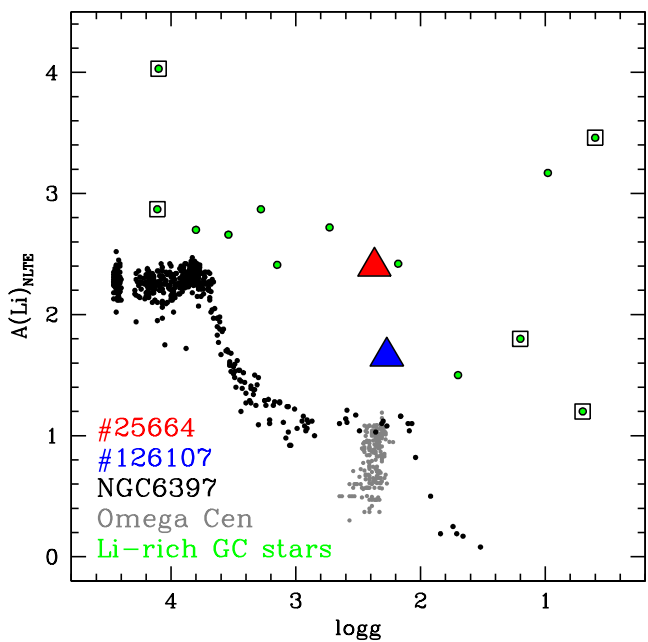}
\caption{Behaviour of A(Li) as a function of log~g for the stars 
in Omega Centauri (grey circles, M2018), in the two Li-rich stars discussed here (red and blue triangles), 
in the Li-rich stars discovered so far in GCs (green circles, squared symbols are the stars for which [Na/Fe] has been measured) and 
the GC NGC~6397 \citep[black circles,][]{lind09}. }
\label{summa}
\end{figure}

We have discovered the first two Li-rich stars in Omega Centauri, 
one of them (\#25664) with an unexpected and exceptional enhancement of Na, 
while the other (\#126107) with a Na abundance compatible with the [Na/Fe] distribution 
of the stars of Omega Centauri with similar metallicity.
Li-rich stars are rare in GCs, only 13 are known so far 
\citep{carney98,kraft99,smith99,ruchti11,koch11,monaco12,dorazi15b,kirby16}, 
10 of them belonging to the RGB. 
Fig.~\ref{summa} summarises the state-of-art about Li-rich GC stars, showing 
A(Li) as a function of log~g (as a proxy of the evolutionary stage) for 
the LRGB stars of Omega Centauri (grey circles), the two Li-rich stars (red and blue triangles) 
and the Li-rich stars discovered so far in GCs (green circles). As a reference 
for the predicted evolution of A(Li), we show 
also A(Li) for the stars in the GC NGC~6397 \citep[black points,][]{lind09}. 
Among the Li-rich stars in GCs, only for 5 stars Na abundances have been measured (squared green circles 
in Fig.~\ref{summa}); all of them have [Na/Fe] compatible with the Na distribution 
of the parent cluster, at variance with \#25664.

Taking into account the effect of the Li depletion due to the first dredge-up
\citep[according to the prescriptions by][]{m12}, 
the two Li-rich giants should have an initial A(Li) of about 3.8 and 3.1 dex, for \#25664 and \#126107, respectively. 
These values are significantly above 
the typical value measured in unevolved Population II stars belonging 
to the {\sl Spite Plateau} (A(Li)$\sim$2.2-2.4 dex) but also higher than the value predicted 
by the Big Bang nucleosynthesis model coupled with the baryon density 
provided by the WMAP and Planck satellites (A(Li)$\sim$2.7 dex). 
In the case of \#25664, its initial A(Li) value is comparable with the 
Li abundance measured in the Li-rich dwarf star discovered in NGC~6397 by \citet{koch11}, 
A(Li)=~4.2 dex,
but the latter has a Na content compatible with the [Na/Fe] 
abundance of 1P stars of NGC~6397 \citep{pasquini14}.
Interestingly, similar lithium enhancements have been recently detected in metal-poor field giant 
stars by \citet{li18}. Also in these cases, abundances of elements other than lithium 
are similar to the trend observed for similar stars.

The origin of Li-rich stars remains still debated and can be ascribed to different processes, 
including external or internal Li production.
One of the most invoked external mechanism to increase A(Li) is the engulfment 
of small bodies, like planets or brown dwarfs
\citep{siess99} that should enhance both Li and Be. 
This scenario seems to be unlikely for the Li-rich stars in Omega Centauri because of the 
low occurrence of metal-poor stars hosting planets \citep[see e.g.][]{j10p}; 
usually the planet engulfment is considered the main mechanism to explain 
Li-rich giants with metallicity higher than $\sim$--0.5 dex \citep{casey16}. 
Also, the engulfment of a planet should also increase the rotational velocity  \citep{siess99} and
the chromospheric activity of the star \citep{fekel93}. We checked that the measured full widths at half maximum are compatible with 
the nominal spectra resolutions and we estimate that in these stars the 
projected rotational velocity is virtually compatible 
with no stellar rotation. 
On the other hand, the Na D lines do not show evidence of circumstellar material.

Another proposed mechanism to explain Li-rich stars but invoking an 
internal production of fresh Li is the Cameron-Fowler mechanism \citep{fowler71}.
In this case we can envisage two possibilities.

The first option is \textit{extra mixing} efficient within the star
during its RGB evolution. This mixing needs to circulate matter between
the base of the convective envelope and a region close to the H-burning
shell. Both speed and depth of the circulation must be such that  $^3He$
from the convective envelope
is transported to temperatures high enough to activate the $^3He(\alpha,\gamma)^7Be$ reaction, with $^7$Be quickly transported back to
cooler regions,
where Li can be then produced by the $^7Be( e^{-},\nu)^7Li$ reaction.
Several papers \citep[see, e.g.][and references therein]{ds95, sb99,
cb00,  dv03, dh04, gpsu09,palmerini11} have discussed in detail scenarios/mechanisms
to achieve this.
\citet{dv03}, in particular, have shown that by calibrating the extra
mixing parameters, enhanced Li abundances in the envelope can be
produced, together with  increased Na (and depleted O). This could for
example go towards explaining the exceptionally high [Na/Fe] measured in
\#25664.

The extra mixing is usually (but not only) associated to rotation, and
the common assumption is that it is activated after the RGB bump, because
the molecular weight barrier between base of the convective envelope and
H-burning shell existing before this event, is expected to inhibit
element transport
between envelope and shell.
Indeed, \citet{cpt05} have shown with full stellar evolution
calculations including rotation and rotational mixings, that this $\mu$
barrier prevents rotational mixing to
be efficient between convective envelope and H-burning shell \citep[see also][]{sm79}.
Our two Li-rich objects are located below the RGB bump, as shown in Figs.~\ref{cmdwfi} and \ref{cmdgaia}.

The second option is production of fresh Li this time in AGB stars \citep[see e.g.][]{ventura10,dantona12},
and mass transfer from a massive star  that evolved through the AGB phase and transferred Li-rich material onto our targets.




From a theoretical point of view, the super-AGB stars, with masses of 
7-8 $M_{\odot}$, may be able to produce simultaneously a large amount of Li and Na \citep{ventura11,dantona12,doherty14}. In these stars,  
Li is produced through the Cameron-Fowler mechanism, while Na is produced 
through the Ne-Na cycle. 
We remind that super-AGB stars could play a relevant role in the explanation
of the MPs observed in all GCs and in Omega Centauri, because they are among 
the candidate polluter stars able to generate the 2P stars.

In this framework, two possible scenarios can be envisaged for \#25664 :\\
{\sl (1)}--{\sl the star formed directly from the gas ejected from super-AGB stars before 
dilution with pristine gas}. Theoretical models of MPs based on AGB stars as main polluters 
need to include some dilution of the AGB ejecta with pristine gas in order to reproduce the observed chemical patterns. 
This is due to the mass dependence of the AGB yields that should lead to a Na-O correlation, 
at variance with what has been observed, if no dilution process is accounted for.
However, it has been proposed that a small fraction of stars may have formed 
from the pure ejecta of super-AGB stars before the dilution process, preserving 
the original chemical composition of these polluting stars \citep{dantona12}.  This mechanism could
explain the presence of a
He-rich (Y$\sim$0.35) sub-population in some systems  
\citep[like Omega Centauri and NGC~2808,][]{norris04,piotto05,sollima05b}, because the super-AGB stars 
should also produce a large amount of He. Hence, in this scenario we expect that \#25664 should have a high He content (Y$>$0.3).
\\ 
{\sl (2)}--{\sl the star was member of a binary system together with a massive star 
and accreted Li-rich material from the companion when the latter reaches the super-AGB phase}. In this case, 
the abundances that we measure are not its original ones but they reflect the chemical composition of the interior of the companion, 
plus some degree of dilution with the convective envelope of the accreting star.
The radial velocities measured from individual FLAMES spectra do not show evidence of variability. 
However, they have been taken on a period of about 3 weeks and we cannot exclude that the star is member of a binary system of a longer period.
Hence, with the current dataset we are not able to disentangle between the two scenarios.

Even if the precise amount of Li and Na produced by AGB stars of different masses 
is highly sensitive to several physical assumptions (i.e. the treatment 
of the convection, overshooting and mass loss), the measured Li and Na abundances of \#25664 are 
qualitatively compatible with those foreseen for the super-AGB stars 
\citep{dantona12,doherty14}.

This Li-Na-rich star may be a direct evidence of
extra-mixing occurring before the RGB Bump, or the first observed relic of the gas ejected from 
super-AGB stars, demonstrating that these stars can play a role to explain the MPs in 
Omega Centauri. In both cases, a future inventory of the chemistry of this star, 
in particular the elements involved in the proton-capture reactions 
(i.e. He, CNO, Mg, Al and their isotopic ratios), is crucial to understand the origin of this unique object. 
For instance, the measure of the $^{12}$C/$^{13}$C isotopic ratio will confirm or refute the pre-bump nature of this star. If this will be confirmed, the \citet{dv03} model may be discarded leaving the AGB scenario only.
If the chemical composition of \#25776 will confirm the scenario related to super-AGB stars, this star will allow to directly study the chemical composition 
of the gas processed in the interior of the super-AGB stars. 


\begin{acknowledgements}
We thank the anonymous referee for his/her useful comments and suggestions.
LM acknowledges support from "Proyecto interno" of the Universidad Andres Bello.
PB acknowledges financial support from the Scientific Council of Observatoire de Paris and from the action
f\'ed\'eratrice  ``Exploitation Gaia''. 
SV gratefully acknowledges the support provided by Fondecyt reg. n. 1170518.
\end{acknowledgements}

\bibliographystyle{apj}

\begin{appendix}
\section{Effective temperatures}
\begin{figure}
\includegraphics[width=\columnwidth]{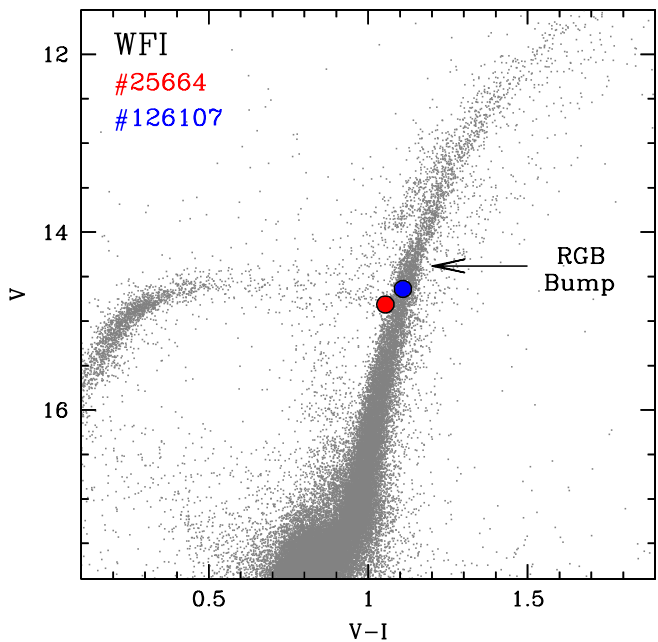}
\caption{(V, V-I) color-magnitude diagram of Omega Centauri 
obtained from the WFI catalog by \citet{bellini09} with marked 
the position of the two Li-rich stars (red circle for \#25664 and 
blue circle \#126107). The arrow indicates the 
luminosity level of the RGB Bump of the metal-poor, dominant population of Omega Centauri.}
\label{cmdwfi}
\end{figure}

\begin{figure}
\includegraphics[width=\columnwidth]{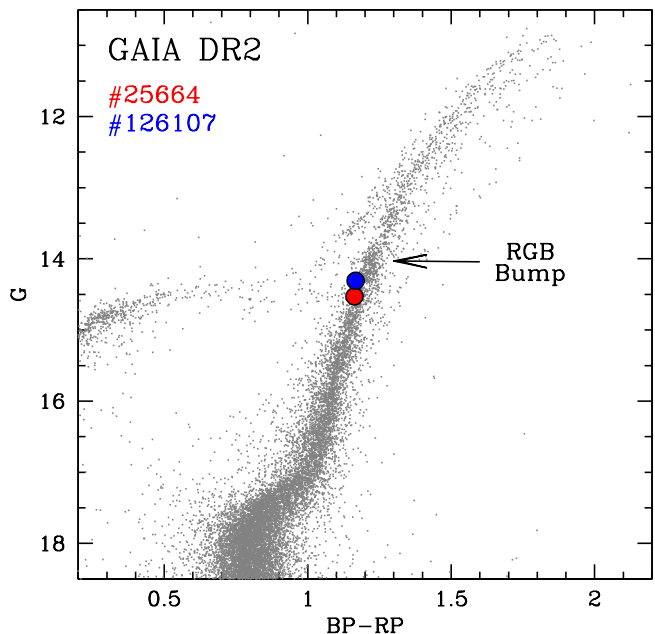}
\caption{(G, BP-RP) color-magnitude diagram of Omega Centauri 
obtained with Gaia Data Release 2 \citep{gaia18} with marked 
the position of the two Li-rich stars (same colors and symbols of Fig.~\ref{cmdwfi}.}
\label{cmdgaia}
\end{figure}

\teff\ is the most crucial parameter in the determination of A(Li), while the other atmospheric parameters have a negligible 
impact on A(Li). 
Fig.~\ref{cmdwfi} and \ref{cmdgaia} show the position 
of the two Li-rich stars in the color-magnitude diagrams  obtained with the 
WFI catalog of \citet{bellini09} and the GAIA Data Release 2 
\citep{gaia18}. In both diagrams the two stars are located 
along the RGB and below the RGB Bump.

\begin{figure}
\includegraphics[width=\columnwidth]{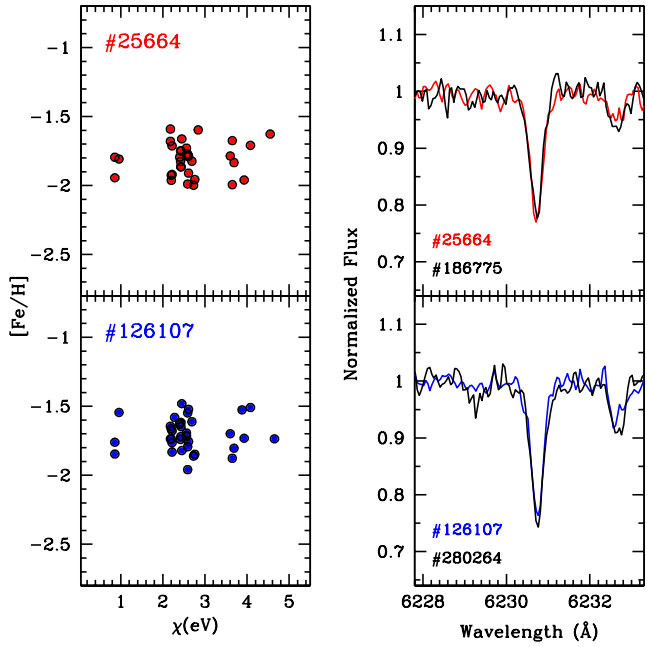}
\caption{{\sl Left panels:} behaviour of the abundance of Fe~I lines as function of the excitation potential $\chi$ for the Li-rich stars, red circles for \#25664 (left upper panel) and blue circles for \#126107 
(left lower panel). 
{\sl Right panels:} spectral region around a Fe~I line in \#25664 (right upper panel, red line) and 
\#126107 (right lower panel, blue line) both compared 
with stars in Omega Centauri with similar metallicity 
and atmospheric parameters and shown in Fig.~\ref{spec}.}
\label{checkteff}
\end{figure}

As a sanity check for the reliability of the photometric 
\teff\ we checked also the excitation equilibrium for the two target stars. Right panels in Fig.~\ref{checkteff} show 
the behaviour of the abundance of neutral iron lines as a function of the excitation potential $\chi$ (red circles for \#25664 and 
blue circles for \#126107). The photometric \teff\ well satisfy 
the excitation equilibrium, confirming the reliability of the 
adopted atmospheric parameters. Also, we show in left panels of Fig.~\ref{checkteff} the strength 
of one Fe~I line in the two Li-rich stars in comparison with 
the same stars already shown in Fig.~\ref{spec}. The very 
similar line strength confirms the similar metallicity and 
atmospheric parameters of the two pairs of stars.

\section{Planet engulfment}
We estimated the amount of rocky material that 
should be ingested by the two stars in order to 
reproduce the measured A(Li), following the same procedure adopted by \citet{mott17}. Assuming that 
two stars have a total mass of 0.8 $M_{\odot}$ each, 
and a mass of the convective envelope of about 0.45 
$M_{\odot}$, the measured A(Li) correspond 
to a total lithium mass of 1.2$\cdot 10^{21}$ Kg 
and 2.1$\cdot 10^{20}$ Kg, for \#25664 and \#126107, respectively. Considering the fractional lithium abundance of rocky material provided by \citet{mcdonough01}, the measured A(Li) could be explained by the engulfment of $\sim$123 and $\sim$22 
Earth masses, for \#25664 and \#126107, respectively. 
For the latter star, we cannot totally rule out that the measured A(Li) can be explained by the engulfment of small bodies.
On the other hand, for \#25664 the engulfment of the expected amount of rocky material ($\sim$123 Earth masses) seems to be unlikely.

\end{appendix}

\end{document}